\documentclass{article}
\usepackage{amssymb}
\usepackage{amsmath}
\usepackage{bm}
\usepackage{graphicx}
\usepackage{color}
\usepackage{authblk}

\begin{document}

\title{Elasticity versus phase field driven motion in the phase field crystal
model}

\author{Amit Acharya$^{1}$, Luiza Angheluta$^{2}$, and Jorge
Vi\~nals$^{3}$}
\affil{$^{1}$Department of Civil and Environmental Engineering, and Center for
Nonlinear Analysis, Carnegie Mellon University, Pittsburgh, Pennsylvania 15213,
USA. \\$^{2}$PoreLab, Njord Centre, Department of Physics, University of Oslo,
P.O. Box 1048, 0316 Oslo, Norway.\\$^{3}$School of Physics and Astronomy,
University of Minnesota, Minneapolis, MN 55455, USA.}

\maketitle

\begin{abstract}
The inherent inconsistency in identifying the phase field in the phase field crystal Theory with the material mass and, simultaneously, with material distortion is discussed. In its current implementation, elastic relaxation in the phase field crystal occurs on a diffusive time scale through a dissipative permeation mode. The very same phase field distortion that is included in solid elasticity drives diffusive motion, resulting in a non physical relaxation of the phase field crystal. We present two alternative theories to remedy this shortcoming. In the first case, it is assumed that the phase field only determines the incompatible part of the elastic distortion, and therefore one is free to specify an additional compatible distortion so as to satisfy mechanical equilibrium at all times (in the quasi static limit). A numerical solution of the new model for the case of a dislocation dipole shows that, unlike the classical phase field crystal model, it can account for the known law of relative motion of the two dislocations in the dipole. The physical origin of the compatible strain in this new theory remains to be specified. Therefore, a second theory is presented in which an explicit coupling between independent distortion and phase field accounts for the time dependence of the relaxation of fluctuations in both. Preliminary details of its implementation are also given.
\end{abstract}

\section{Introduction}
The phase field crystal (PFC) model has become a successful and widely used description of the mesoscale response of a nonequilibrium crystalline phase \cite{re:elder02}. Since, by construction, the need to resolve the time scale associated with lattice vibration is eliminated, its computational implementation can efficiently access long time phenomena that are difficult to describe by molecular dynamics simulation \cite{re:emmerich12,re:momeni18}. One of the strengths of the model is the deceivingly simple description of defected solids. A longstanding debate, to which we wish to contribute here, is the nature of the phase field itself, and the proper long wavelength description of a defected elastic material.

A widely used interpretation is that the phase field represents the mass density of the solid \cite{re:stefanovic06,re:heinonen16}. In fact, phase field crystal model free energies have been derived by using Density Functional Theory methods, with the expectation of obtaining functionals of the mass density as the single, relevant order parameter that describe the structure and elastic response of a crystalline solid in equilibrium \cite{re:elder07,re:huang10}. In addition, the free energies obtained have been shown to provide a reasonable description of the freezing phase transition \cite{re:archer19}. However, extensions so that the set of slow variables in the Density Functional Theory also includes the momentum density have not been considered to date (except for colloidal systems \cite{re:archer09}). Therefore, the constitutive laws relating momentum currents (stresses) and their conjugate strains, as well as their coupling to the phase field, remain completely phenomenological.

Most phase field crystal model implementations begin with the separation of the mass density into two components: A slowly varying part in space $\rho$ -- which is then identified with the standard hydrodynamic mass density -- and a second component, $\psi$, which is obtained from the location of the Bragg peaks of the crystalline structure under consideration \cite{re:heinonen16}, and that describes crystalline order. The mass density is decomposed as,
$$
\rho(\mathbf{x}) + \psi(\mathbf{x}) = \rho(\mathbf{x}) + \sum_{n} A_{n} e^{i \mathbf{q}^{(n)} \cdot \mathbf{x}}.
$$
where the $A_{n} = |A_{n}| e^{i \theta_{n}}$ are slowly varying complex amplitudes. The phase $\theta_{n}$ represents a lattice distortion $\theta_{n} = - i \mathbf{q}^{(n)} \cdot \mathbf{u} (\mathbf{x},t))$. The vectors $\mathbf{q}^{(n)}$ are the critical eigenmodes of the model's free energy at the bifurcation point to the modulated phase, and $\mathbf{u} (\mathbf{x},t)$ is the displacement field \cite{re:stefanovic06,re:heinonen16,re:skaugen18}. Although this relation is only expected to hold for weakly distorted lattices, it is commonly used to study defected configurations as well. It is one of the goals of this paper to discuss the limitations inherent in this mapping between the phase field $\psi$ and the displacement field $\mathbf{u}$.

The phase field crystal is one among several theories that allow a short scale regularization of defect core divergences inherent in classical elasticity, while allowing for the treatment of defect topology and motion, grain boundary energetics and motion, and explicit expressions for the associated mobilities. For equilibrium studies, the only constitutive input required is the (defect free) equilibrium free energy, a non convex functional of the phase field, with a minimizer that corresponds to a spatially periodic configuration. For nonequilibrium problems, current implementations of the phase field crystal generally assume gradient flows driving the evolution of a configuration towards a local minimizer of the free energy functional. Importantly for the regularization, the phase field is smooth everywhere, including in defected configurations, but its phase is singular at defect cores. The proper treatment of this singularity is the main focus of this paper.

\section{Thermodynamics and Linear Response}
\label{sec:thermodynamics}

The defining feature of a solid material subject to a small distortion is its elastic response under the application of an external body force $\mathbf{f}_{ext}(\mathbf{x})$. Within linear elasticity, the energy of the distortion is
\begin{equation}
  {\cal F} = \frac{1}{2} \int d^{d}x \; C_{ijkl} \bm{\epsilon}_{ij} \bm{\epsilon}_{kl} - \int d^{d}x \; \mathbf{f}_{ext} \cdot \mathbf{u} \quad{\rm with,} \quad \bm{\epsilon}_{ij} = \frac{1}{2} \left( U_{ij} + U_{ji}\right)
\end{equation}
with the distortion tensor $U_{ij} = \partial_{i} u_{j}$, and $\mathbf{u}$ the displacement vector, with the usual definitions of strain $\epsilon_{ij}$ and elastic constants $C_{ijkl}$. In contrast, subjecting a phase field crystal to strain involves an indirect transformation, as it involves imposing an externally fixed chemical potential $\mu_{ext} $, and requiring that the distorted equilibrium configuration minimizes ${\cal F} = {\cal F}_{sh} - \int d^{d}x \; \mu_{ext}(\mathbf{x}) \psi(\mathbf{x})$, where ${\cal F}_{sh}$ is typically chosen to be of the Swift-Hohenberg type \cite{re:heinonen14}.  Alternatively, one can directly impose the deformation on the order parameter $\psi$ directly. In neither case is it possible to, say, specify body forces or tractions on the boundary of the phase field crystal.

This is not a shortcoming in equilibrium. Let  $\delta \psi$ be the local variation of the phase field $\psi$ following a small distortion of the equilibrium lattice. Since mass is conserved, this variation is $\delta \psi = - \partial_{k} (\psi \delta u_{k})$, where $u_{k}$ is the $k$-th component of the displacement vector. From this relation, the variation $\delta {\cal F}/\delta u_{k} = \psi \partial_{k} (\delta {\cal F}/\delta \psi)$ follows as the change in free energy due to the distortion. The thermodynamic stress $\mathbf{T}$ is defined from the same energy as $\partial_{j} T_{ij} = - \delta {\cal F}/\delta u_{i}$, and therefore $\partial_{j} T_{ij} = - \psi \partial_{i}(\delta {\cal F}/\delta \psi)$. This relation is correct in equilibrium where both sides of the equation independently vanish. However, this is not the case outside of equilibrium: The right hand side of the equation is but one contribution to the reversible stress, and the equality does not hold in general. Note that the equality between what is an {\em elastic} body force $\partial_{j} T_{ij}$ and a {\em configurational} force $- \psi \partial_{i}(\delta {\cal F}/\delta \psi)$ follows directly from the assumption that lattice displacement and variation of $\psi$ are not independent, but rather related through $\delta \psi = - \partial_{k} (\psi \delta u_{k})$.

In nonequilibrium studies, this difficulty can be remedied, although somewhat artificially, by assuming that the slow component $\rho$ is defined to be the mass density, which is indeed conserved, but that the crystalline component $\psi$ is not, regardless of the fact that both represent a mass density (e.g., Ref. \cite{re:heinonen16}). As a consequence, both fields are now subject to independent variation. Under these conditions, the authors of Ref. \cite{re:heinonen16} find that the mass velocity $\mathbf{v}$ is different than the lattice velocity $\partial_{t} \mathbf{u}$,
$$
\mathbf{v} = \partial_{t} \mathbf{u} + \frac{1}{2 A_{0}^{2}} \mu_{\eta} \frac{\delta {\cal F}}{\delta \mathbf{u}},
$$
where $A_{0}$ is the value of the equilibrium amplitude $A_{n}$ in the hexagonal lattice studied in that reference, and $\mu_{\eta}$ is the mobility associated with the gradient flow of $\psi$. The difference between the two velocities is a dissipative term, and it corresponds to a permeation mode that decouples lattice and mass motion.

Within this decomposition, and with explicit consideration of kinetic energy terms $\frac{1}{2} \int \rho v^{2} d^{d}x$ and momentum conservation, the following dispersion relation for the transverse displacement $\mathbf{u}_{\perp} \cdot \mathbf{q} = 0$ in the overdamped limit results,
\begin{equation}
\omega_{\perp}(q) = - \frac{i}{2} B \mu_{\eta} q^{2} \pm \frac{q}{2} \sqrt{ \frac{B}{\rho}\left(12 A_{0}^{2} - B \mu_{\eta}^{2} \rho q^{2} \right) }.
\label{eq:disprel}
\end{equation}
where $B/2$ is the coefficient of the gradient kernel in the Swift-Hohenberg free energy. This is an interesting result, as the decoupling allows for the emergence of propagating modes at low $q$, not just the purely diffusive response of the classical phase field crystal. This result needs to be contrasted with the classical dispersion relation for transverse modes of an isotropic crystal (with Newtonian viscosity) which is $\omega = -i (\eta/2 \rho) q \pm q \sqrt{\mu/\rho - \eta^{2}/(4 \rho^{2})}$, with $\mu$ the shear modulus, and $\eta$ the shear viscosity. The speed of propagation of transverse distortions is  $c_{\perp} = \sqrt{\mu/\rho}$. The associated transverse susceptibility is $\chi(\mathbf{q},\omega) = \mathbf{u}_{\perp}(\mathbf{q},\omega) /  \mathbf{f}_{\perp,ext}(\mathbf{q},\omega) = (1/\rho) \left( -\omega^{2} + (\mu - i \omega \eta)q^{2}/\rho \right)^{-1}$. At zero frequency, the response is that of an elastic solid $1/(\mu q^{2})$. For the phase field crystal the inverse susceptibility at zero frequency is $\chi^{-1}(\omega = 0) \propto (3B A_{0}^{2}/\rho) q^{2} + {\cal O}(q^{4})$ and hence the phase has a solid response at low frequency with shear modulus proportional to $3B A_{0}^{2}/\rho$. Two features are of interest. First, as $q$ increases, the square root in Eq. (\ref{eq:disprel}) becomes imaginary, and the response of the phase field crystal is purely diffusive. This is relevant for defected solids because gradients of the phase field amplitude are large near cores. Second, the imaginary part of $\omega_{\perp}$ is due entirely to permeation. In a conventional elastic solid, the imaginary part is the shear kinematic viscosity, which is essentially zero for practical purposes.  Hence an elastic solid remains distorted under tension, and it can store elastic energy. In the phase field crystal, on the other hand, the ratio of the imaginary to the real part of the susceptibility is  $\omega (\rho \mu_{\eta} q)/(3 \phi^{2})$. The denominator is small (on the order of the dimensionless distance to the ordering transition for the phase field crystal), and hence permeation becomes negligible only for very small values of the mobility $\mu_{\eta}$, or as $q \rightarrow 0$. Making the mobility very small is tantamount to eliminating dissipative relaxation of $\psi$, the raison d'etre of the model in the first place. Alternatively, elastic response is recovered only in the limit $q \rightarrow 0$ which is not appropriate for the study of elastic solids. Therefore, whereas an elastic solid reversibly stores elastic energy under traction, the response of the phase field crystal is primarily diffusive. Ultimately, the elastic response and the permeation mode in the phase field crystal both result from the curvature of the planes of constant $\psi$, both proportional to the transverse diffusion coefficient, which is proportional to the same coefficient $B$. Although this difficulty has been recognized for a long time, and a number of modified models have been introduced to allow for relaxation of the phase field in a time scale faster than diffusion \cite{re:stefanovic06,re:majaniemi07,re:heinonen14,re:heinonen16,re:zhou19}, it still remains an open question.

Finally, as concerns the dissipative mechanical response of crystalline solids, it must be recognized that the phenomenon of plasticity is strongly dissipative (with greater than $50 \%$ of applied power being converted to heat, more often close to $90 \%$) and this dissipation arises primarily due to energy taken up in incoherent atomic vibrations produced due to the motion of crystal defects through the body.

\section{Defected phase field crystal}

The difficulties just described become more serious when the crystal is allowed to have topological singularities. In this case, the permeation mode just described acquires topological content. Let $\alpha_{ij}$ be the dislocation density tensor defined so that its integral over a surface is the sum of the Burgers vectors of the dislocation lines piercing the surface \cite{re:kosevich79}. For an isolated dislocation line it can be written as $\alpha_{ij} = t_{i} b_{j} \delta(\bm{\zeta})$, where $\mathbf{t}$ is the unit tangent to the line, $\mathbf{b}$ is the Burgers vector of the line, and $\bm{\zeta}$ is the two dimensional radial coordinate (with origin at the line) on the plane normal to the line tangent. The tensor $\bm{\alpha}$ describes the incompatibility of the lattice distortion as
\begin{equation}
  \alpha_{ik} = - \epsilon_{ilm} \partial_{l} U_{mk},
  \label{eq:alpha_def}
\end{equation}
where $\bm{\epsilon}$ here is the antisymmetric tensor. An immediate consequence of this result is that even though the phase field $\psi$ is regular everywhere, since its phase $\theta_{n}$ is not, the displacement field is not uniquely determined by a configuration of the phase field $\psi$, as was given, for example, in the relation between the phase of $\psi$ and the displacement in the previous section. Furthermore, if one were to assume that $\mathbf{v} = \partial_{t} \mathbf{u}$, or that $\partial_{i} v_{k} = \partial_{t} U_{ik}$, then simply by taking the time derivative of Eq. (\ref{eq:alpha_def}), one has $\partial_{t} \bm{\alpha} = 0$, the dislocation line does not move. Dislocation motion requires the explicit uncoupling of mass and lattice motion as in,
\begin{equation}
\partial_{i} v_{k} = \partial_{t} U_{ik} - {\cal J}_{ik}.
\label{eq:uncoupling}
\end{equation}
Substitution into the time derivative of Eq. (\ref{eq:alpha_def}) yields,
\begin{equation}
  \partial_{t} \alpha_{ik} + \epsilon_{ilm} \partial_{l} {\cal J}_{mk} = 0,
\end{equation}
which is precisely the statement of conservation of topological charge under the motion of the dislocation line. This result is standard in continuum dislocation theory \cite{mura1963continuous, re:kosevich79} (for extensions to finite deformations, see \cite{fox1966continuum}, \cite[App.~B]{acharya2011microcanonical}). The dislocation charge density current can be written as as ${\cal J}_{ik} = \epsilon_{irs} \alpha_{rk}(v_{s} + V_s)$, where $\mathbf{V}$ is the velocity of the dislocation line relative to the material velocity $\mathbf{v}$ \cite{re:acharya07}. Unlike the purely dissipative character of the permeation mode of Sec. \ref{sec:thermodynamics}, the dislocation density current has both reversible and dissipative contributions. 

In physical terms, it remains to be resolved whether the phase field crystal model (an intrinsically dissipative theory) can approximate to some degree both reversible topological charge density currents as well as the local dissipative processes around dislocation cores during line motion, including therefore Peierls stresses and barriers. The phase field crystal gives an explicit expression for ${\cal J}_{mk}$ in terms of the time rate of change of the phase field amplitude \cite{re:skogvoll22}. 
%
%
To the extent that in this latter treatment the evolution of the phase field is purely dissipative, it follows that the defect current generated by the motion of defects is a source of energy dissipation. 

\subsection{Decoupling through a compatible distortion}
\label{sec:compat}

One first possible approach at making the phase field $\psi$ independent from distortion is given in Ref. \cite{re:skaugen18b}. Given that the phase field crystal can describe topological singularities and their motion, but not elasticity, it seems reasonable to use $\psi$ to determine the incompatible part of the distortion only, and then introduce a new, independent, compatible distortion. The former relaxes in plastic time scales as $\psi$ evolves diffusively, whereas the second relaxes quickly and accounts for elastic response.

\begin{figure}
  \includegraphics[width=0.85\linewidth]{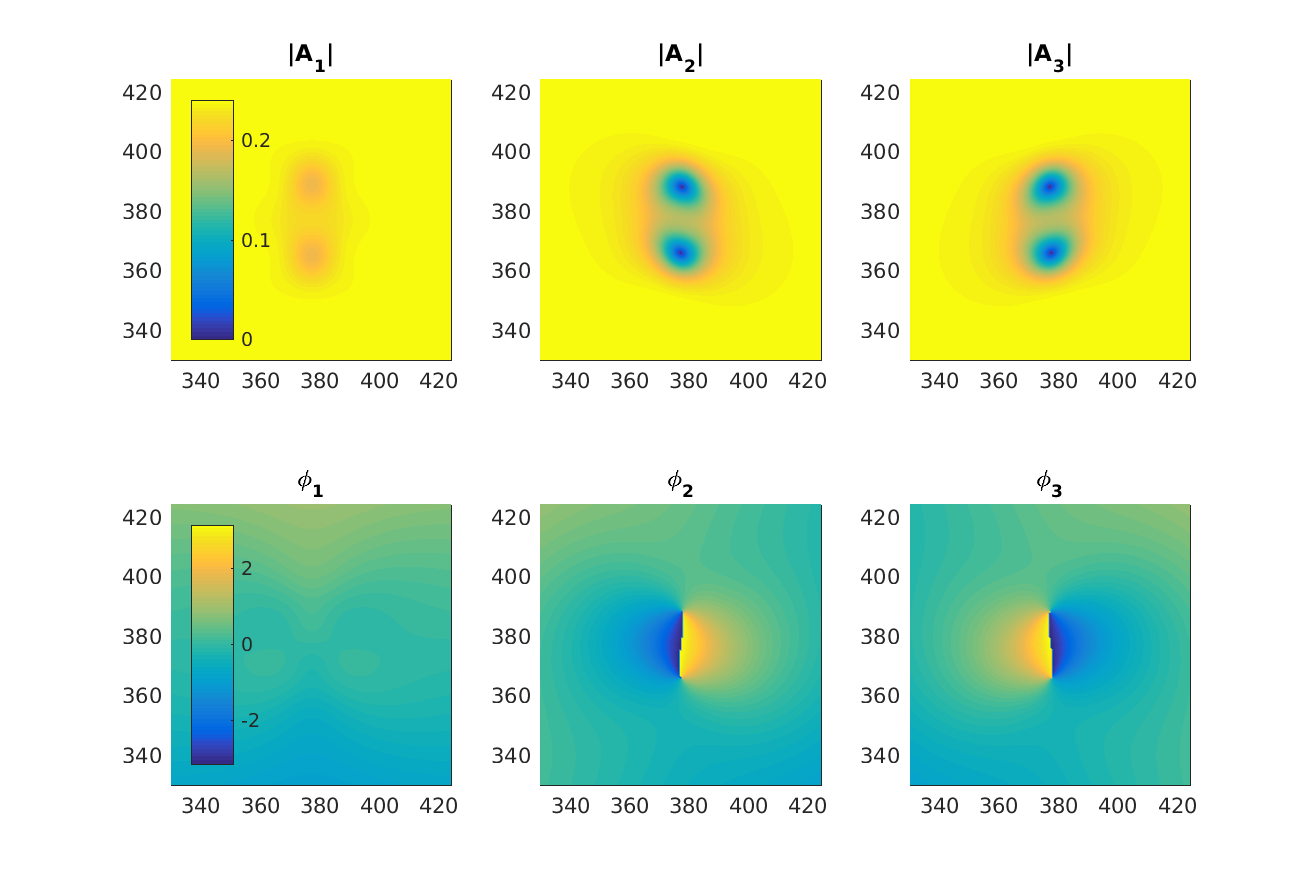}
  \caption{Amplitudes and phases of the phase field $\psi$ for a dislocation dipole in a two dimensional hexagonal lattice. Only two of the three wave amplitudes of the order parameter go to zero at the defect core. The corresponding phases exhibits a discontinuity along a line joining the cores. The asymmetry in the amplitude contours reflects the different orientation of the two base waves which are rotated by $120^{o}$ relative to each other.}
  \label{fi:singularities}
\end{figure}

While the phase field is regular at defects, its phase is singular, fact that can be used to define the Burgers vector density in two dimensions, or the dislocation density tensor in three dimensions. In two dimensions, one has
$$
\oint d \theta_{n} = q_{j}^{(n)} b_{j} 
$$
where the Burgers vector is defined as $\oint d \mathbf{u} = - \mathbf{b}$ (Fig. \ref{fi:singularities}, see Ref. \cite{re:skaugen18} for details). 
The singular phases can be used to directly compute a configurational distortion tensor, as will be described in Sec. \ref{sec:Qtensor} (see also Ref. \cite{re:acharya20}).

Since the phase field crystal satisfies its own evolution equation, it is not necessary to compute the incompatible distortion explicitly. However, the compatible part must be determined, and one can do so by using the requirement that the crystal is in mechanical equilibrium at all times (which is reasonable in the diffusive time scale of dislocation motion determined by the evolution of $\psi$). In order to determine the compatible distortion, it is possible, within linear elasticity, to use the incompatible part as the source of stress. A configurational stress is first defined, which is regular even at dislocation cores. Consider the change in phase field free energy $\Delta {\cal F} = {\cal F}[\psi(\mathbf{x}')] - {\cal F}[\psi(\mathbf{x})]$ due to a small affine distortion $\mathbf{x}' = \mathbf{x} + \mathbf{u}(\mathbf{x})$. Then a configurational stress $T_{ij}^{\psi} = \langle \frac{\partial f}{\partial(\partial_{i} u _{j})} \rangle$ is defined where $f$ is the free energy density, and $\langle . \rangle$ stands for an average over the unit cell of the phase field crystal in order to remove spatially oscillatory components. As is well known, $\nabla \cdot T_{ij}^{\psi} \neq 0$ in general. The method of Ref. \cite{re:skaugen18b} aims at introducing an independent (and compatible) distortion $\mathbf{U}^{\delta}$, so that the combined corresponding stresses satisfy $\nabla \cdot (\mathbf{T}^{\psi} + \mathbf{T}^{\delta}) = 0$. By construction, this distortion is compatible, and can be integrated to find the corresponding displacement $\mathbf{u}^{\delta}$ everywhere.
The compatible displacement is computed therefore from $\nabla \cdot ({\cal C} : \nabla \mathbf{u}^{\delta}) = - \nabla \cdot \mathbf{T}^{\psi}$ where ${\cal C}$ is the tensor of elastic constants. In the case of a two dimensional hexagonal lattice (elastically isotropic) considered in Ref. \cite{re:skaugen18b}, the total stress $\mathbf{T} = \mathbf{T}^{\psi} + \mathbf{T}^{\delta}$ was expressed in terms of the Airy stress function $\chi$ as $T_{ij} = \epsilon_{ik} \epsilon_{jl} \partial_{k} \partial_{l} \chi$. Then,
\begin{equation}
  (1 - \frac{\lambda}{2(\lambda + \mu)} ) \nabla^{4} \chi = \epsilon_{ik} \epsilon_{jl} \partial_{i} \partial_{j} T_{kl}^{\psi} - \frac{\lambda}{2(\lambda + \mu)} \nabla^{2} T_{kk}^{\psi} ,
\label{eq:elastic_eq}
\end{equation}
in terms of the Lame coefficients of the lattice. Given an instantaneous configuration of $\psi$, the right hand side of Eq. (\ref{eq:elastic_eq}) is known. Solving for $\chi$, one then has
\begin{equation}
  T_{ij}^{\delta} = \epsilon_{ik} \epsilon_{jl} \partial_{k} \partial_{l} \chi - T_{ij}^{\psi},
\end{equation}
and, from it, the corresponding distortion $\mathbf{U}^{\delta}$. 

It is the case then that if a fixed configuration of the phase field is transformed as $\psi(\mathbf{x}) \rightarrow \psi(\mathbf{x}+\mathbf{u}^{\delta})$, then the new stress $\mathbf{T}^{\psi}$ is divergenceless. Furthermore, since the distortion $\mathbf{u}^{\delta}$ is compatible, this transformation does not change the topological content of the field $\psi$. The temporal evolution of the phase field is assumed in Ref. \cite{re:skaugen18b} to include the dissipative evolution given by the gradient flow induced by ${\cal F}$, plus and adiabatic distortion by $\mathbf{u}^{\delta}$ at each time, so that the combined motion takes place on the restricted manifold defined by $\nabla \cdot \mathbf{T}^{\psi} = 0 $. Clearly, this evolution is arbitrary, and unphysical. It fails to describe the physical mechanism by which the phase field relaxes in a fast time scale to reach mechanical equilibrium for a given, fixed, distribution of defects. In this model, the relaxation is infinitely fast and accomplished by a simple displacement on $\psi$.

\begin{figure}
  \includegraphics[width=0.85\linewidth]{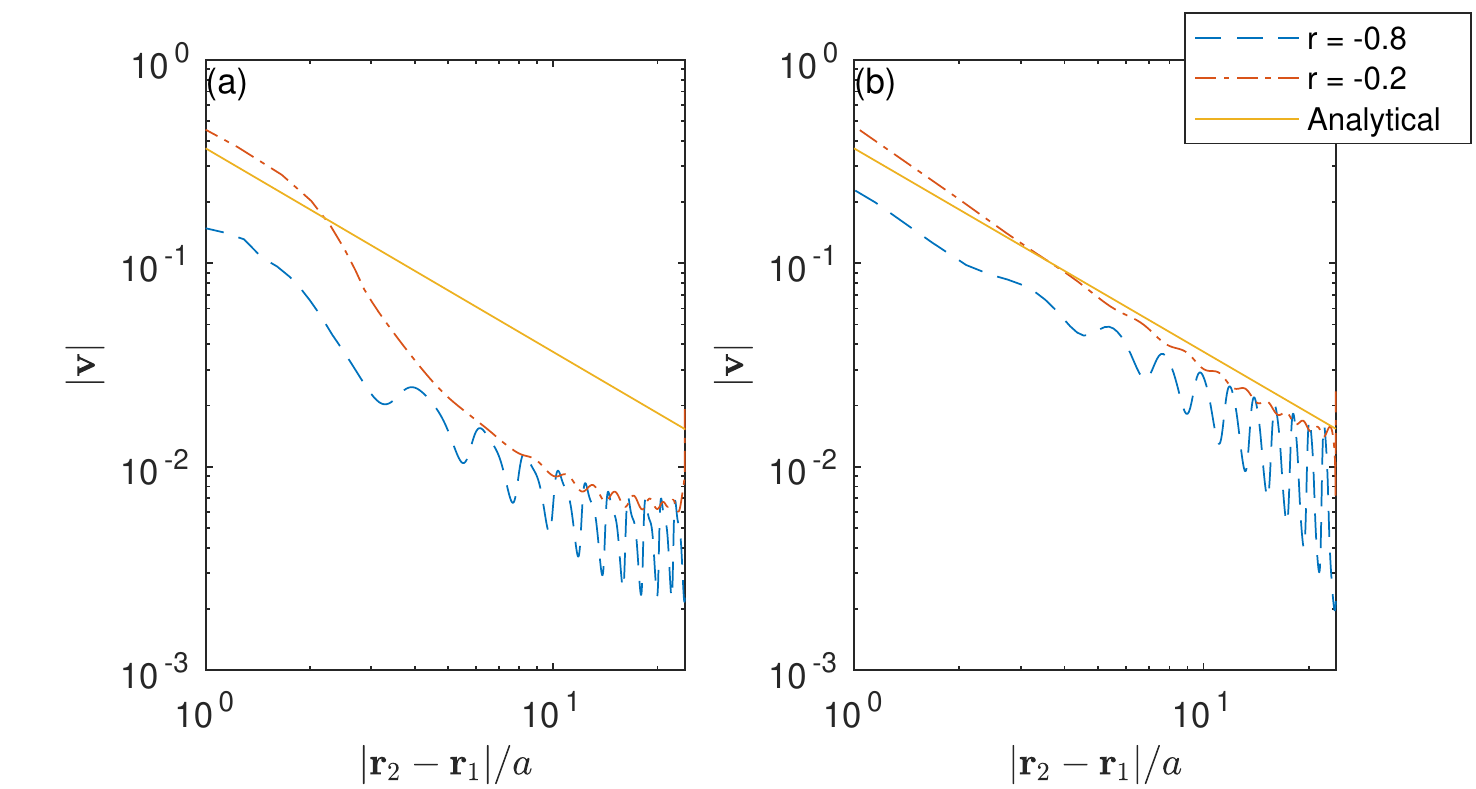}
  \caption{Relative velocity of two edge dislocations of opposite Burgers vector in a hexagonal lattice as a function of their dimensionless separation. Left: evolution given by a purely dissipative phase field crystal. The parameter $r$ in the box equals the coefficient of the quadratic term in the phase field free energy. Right: evolution given by the model described in this section \cite{re:skaugen18}. Oscillations for large values of $r$ in both cases (far from the bifurcation threshold to the hexagonal phase) correspond to transient pinning effects described earlier \cite{re:boyer02}.}
  \label{fi:velocity}
\end{figure}

The resulting defect core motion observed deviates significantly from the results given by the classical phase field crystal method (Fig. \ref{fi:velocity}). Similar conclusions were already reported in Ref. \cite{re:berry06} that noted that different methods of lattice distortion application led to different dislocation motion. Two limiting cases were considered: a uniform, affine distortion of the reference lattice (termed \lq\lq rigid displacement"), and a \lq\lq relaxational method" in which a thin layer of boundary sites were displaced, and the rest of the lattice allowed to relax under phase field diffusion. The motion of a dislocation in the same two dimensional, hexagonal lattice studied here showed a substantial dependence on the method of lattice distortion used.


\subsection{The phase field as a configuration tensor}
\label{sec:Qtensor}

A different scheme in which to treat the distortions of the phase field crystal and of the underlying material as independent has been given in Ref. \cite{re:acharya20}. It combines the Field Dislocation Mechanics model (FDM, cf. \cite{roy2005finite, zhang2015single,arora2020finite,acharya2022action}) -- demonstrated to represent features like large `stress-free' shear bands in the presence of significant material inertia, sub-inter-supersonic dislocation motion in accord with molecular dynamic simulations in \cite{gumbsch1999dislocations}, dislocation annihilation, dissociation, dipole and pile-up interactions and equilibria, and Peierls stress (all without any parameter adjustment) -- with the phase field crystal methodology. The strength of the FDM methodology is in providing a framework for calculating (nonlinear) elastic fields in bodies of arbitrary shape and elastic anisotropy, and for evolution accounting for the topological charge conservation of dislocations. Suitably interpreted, it provides flexibility in the representation of lattice periodicity in a continuum pde model. The work in \cite{re:zhang16} utilized Generalized Stacking fault energies (GSFs) (determinable from interatomic potentials \cite{vitek1968intrinsic}, as also utilized in phase Field methods, e.g. \cite{shen2003phase}), a building block analogous to the choice of a free energy functional from the Swift-Hohenberg family in the phase field crystal. A goal of \cite{re:acharya20} is to use Swift-Hohenberg type models instead of GSFs.

In Ref. \cite{re:acharya20}, the phase field crystal is used to define a configurational distortion tensor $\mathbf{P}$, independent of the material distortion, and that is well defined in a defected medium. Given a general coordinate transformation defined as the following transformation of the basis vectors, $\overline{e}_{i} = P^{m}_{.\;i} e_{m}$, the induced transformation in reciprocal space is
$$
q^{i} = (P^{-1 \ T})^{i}_{.j} \ \overline{q}^{\ j}
$$
One can use this result to define a configurational distortion tensor from the local phase gradients of the phase field (the distorted configuration), and a the reciprocal basis of the equilibrium, undistorted configuration,
\begin{equation}
\mathbf{q}_{0}^{i} = P^{-1 \ T} \nabla \theta_{i}.
\label{eq:pdef}
\end{equation}
The configurational distortion tensor $\mathbf{P}$ is a point wise functional of the phase field $\psi$, which is defined so as to coincide with the inverse elastic distortion of the medium $\mathbf{W}$ only in equilibrium. Away from equilibrium, one allows relative fluctuations between the two so that the elastic response of the medium is captured by $\mathbf{W}$ whereas the diffusive relaxation of the phase field is described by $\mathbf{P}$.

In order to further motivate this definition, it is useful to consider the small deformation limit. Note that the authors of Ref. \cite{re:acharya20} use a different sign convention for the Burgers vector $\oint d \mathbf{u} = \mathbf{b}$, and also the transpose of the distortion above ($U_{ij} = \partial_j u_i$ for a compatible distortion). The inverse elastic distortion satisfies $\mathbf{W} = \mathbf{I} - \mathbf{U}$, where $\mathbf{U}$ is the fundamental measure of elastic distortion. Analogously, one defines for small deformations $\mathbf{Q} = \mathbf{I} - \mathbf{P}$. Now if the phase of the order parameter is tied to the distortion, $\theta_{n} = \mathbf{q}^{(n)} \cdot \mathbf{r} - \mathbf{q}^{(n)} \cdot \mathbf{u}$, and therefore $\partial_{j} \theta_{n} = q^{(n)}_{j} - q^{(n)}_{i} \partial_{j} u_{i}  = q^{(n)}_{j} - q^{(n)}_{i} U_{ij} = q^{(n)}_{i}(\delta_{ij} - U_{ij})$. Inverting it to put in the form of Eq. (\ref{eq:pdef}) one finds $P_{ij} = \delta_{ij} - U_{ij}$. Therefore $\mathbf{Q} = \mathbf{U}$ in the case in which the phase of the phase field can be expressed as indicated. However, they are treated in the theory as independent. The theory considers an elastic energy functional of the distortion, a phase field free energy of the Swift-Hohenberg type, and a penalty term which is a functional of relative fluctuations between the elastic distortion $\mathbf{U}$ and the configurational distortion $\mathbf{Q}$,
\begin{equation}
  {\cal F}(\mathbf{U},\psi) = {\cal F}_{el}(\mathbf{U}) + C_{sh} {\cal F}_{sh}(\psi) + \frac{C_{w}}{2} \int d^{d} x \; (\mathbf{U} - \mathbf{Q})_{sym}^{2} 
  \label{eq:fe}
\end{equation}
where $C_{sh}$ and $C_{w}$ are two coupling constants. In the governing equations for the nonequilibrium evolution of the fields to be shown below, the constant $C_{w}$ corresponds to an inverse time scale over which the configurational distortion and the elastic distortion relax towards each other.

An initially defected configuration will be described by an order parameter field $\psi$ so that topological defects will be located in regions of non zero curl of $\mathbf{Q}$, defined by the point wise oriented triad in reciprocal space, generally not orthonormal, as given.  If $C_{w}$ and the scale of ${\cal F}_{sh}$, $C_{sh}$, are simultaneously large, the phase field will relax quickly (and diffusively) to a local minimum of the constrained free energy, relatively independently of the changes in elastic energy that the relaxation might impose. This process will be accompanied by a relaxation of the elastic distortion in phonon time scales, very quickly if the quasistatic limit is invoked. Further evolution will be slow, driven by a Peach-Kohler force resulting from both mechanical stress and phase field configurational distortion (\ref{eq:PK}).

Given the free energy in Eq. (\ref{eq:fe}), a (mechanical) dissipation inequality is introduced in Ref. \cite{re:acharya20} by requiring that the external power supplied to the body through the applied tractions minus the rate of change of the free energy and kinetic energy be non-negative, which reduces to the statement
\begin{equation}
  \int d^{d} x \mathbf{T} : \nabla \mathbf{v} \ge \frac{d {\cal F}}{dt},
\end{equation}
assuming balance of linear momentum is satisfied. Further, one also introduces the fundamental kinematic relation of Field Dislocation Mechanics theory 
\begin{equation}\label{eq:el_distort}
    \dot{\bm{U}} = \nabla \mathbf{v} - \bm{\alpha} \times \mathbf{V},
\end{equation}
where $\mathbf{v}$ is the material velocity, $\bm{\alpha} = \mbox{curl}\, \mathbf{U}$ and $\mathbf{V}$ is the dislocation velocity (compare with Eq. (\ref{eq:uncoupling}), with a change of sign in $\bm{\alpha}$ due to the sign convention in the definition of the Burgers vector). The dissipation inequality and the kinematic constraint lead to constitutive equations for the fields. The reversible part of the stress is given by
\begin{equation}
    \mathbf{T} = \partial {\cal F}_{el}/\partial \bm{\epsilon} 
    + C_w (\mathbf{U} - \mathbf{Q})_{sym}.
\end{equation}
The evolution equation for the phase field is given by
\begin{equation}\label{eq:S-H}
  \dot {\psi} = - L \left[ \frac{\delta {\cal F}_{sh}}{\delta \psi} + \frac{C_{w}}{2} \frac{\delta}{\delta \psi} \int d^{d} x \; (\mathbf{U} - \mathbf{Q})_{sym}^{2}\right]
  ,
\end{equation}
where $\dot{\psi}$ is the material time derivative of $\psi$ and $L$ is a kinetic mobility. The variational derivative in the right hand side of Eq. (\ref{eq:S-H}) can be evaluated explicitly from the phase field as follows. We begin by using the result \cite{re:skogvoll22}
$$
\sum_{n} q_{0\;i}^{(n)} q_{0\;j}^{(n)} = \frac{N q_{0}^{2}}{d} \delta_{ij}
$$
where $N$ is a constant that depends on the lattice symmetry, $d$ is the space dimension \cite{re:skogvoll22}, and, as before, $q_{0} = 1$. For example, $N=3$ for a two dimensional hexagonal lattice, or $N=12$ for a three dimensional bcc lattice. Therefore, the configurational distortion tensor is given by,
$$
P_{ij} = \frac{d}{N} \sum_{n}  q_{0\;j}^{(n)}   \partial_{i} \theta
$$
Since
$$
\nabla \theta_{n} = \frac{1}{|A_{n}|^{2}} {\rm Im}(A_{n}^{*} \nabla A_{n}),
$$
then ($Q = I - P$)
$$
Q_{ij} = - \frac{d}{N} \sum_{n} \frac{q_{0\;j}^{(n)}}{|A_{n}|^{2}} {\rm Im}(A_{n}^{*} \nabla A_{n}).
$$
This is the relationship between the configurational distortion and the amplitudes of the phase field. Finally, the required variational derivative follows as,
$$
\delta {\cal I} = \delta \frac{1}{2} \int d \mathbf{x} \; \left( U - Q \right)^{2} = - \sum_{n} \int d \mathbf{x} \; \left[ \frac{\delta I}{\delta A_{n}} \delta A_{n} +  \frac{\delta I}{\delta A_{n}^{*}} \delta A_{n}^{*} \right].
$$
with,
$$
\frac{\delta I}{\delta A_{n}} = (U_{ij}-Q_{ij}) \frac{\partial Q_{ij}}{\partial A_{n}} - \partial_{k} \left[ (U_{ij}-Q_{ij}) \frac{\partial Q_{ij}}{\partial (\partial_{k} A_{n})} \right].
$$

The set of evolution statements is completed by balance of linear momentum that controls the evolution of the material velocity (of course coupled to \eqref{eq:el_distort} and \eqref{eq:S-H}):
\begin{equation*}
    \rho \, \dot{\mathbf{v}} = \mbox{div}\,  \mathbf{T}
\end{equation*}
or, in the quasi-static case,
\begin{equation*}
    \mbox{div}\, \dot{\mathbf{T}} = \mathbf{0};
\end{equation*}
in the latter case, the equation for the velocity is akin to Stokes flow with source terms. 
Equation (\ref{eq:el_distort}) is the statement of the partition of the (compatible) material velocity gradient into the elastic strain rate and the plastic strain rate, the latter arising from the transport of dislocations; in geometrically exact finite deformation analysis, this is a consequence of Burgers vector conservation (with a minimalistic assumption of setting a free gradient to zero). This framework allows large permanent deformations with minimal elastic energy cost, encoding the plastic strain produced due to dislocation motion, to be an outcome of the theory.

Finally, we mention that an expression for the Peach-Kohler force also follows directly from the dissipation inequality \cite{re:acharya20},
\begin{equation}
   \mathbf{V} = \mathbf{M}:\mathbf{X}\left(\mathbf{T}^T \bm{\alpha}\right); \qquad V_{i} = M_{ij} \epsilon_{jrk}  T_{sr}
    \alpha_{sk},
    \label{eq:PK}
\end{equation}
where $\mathbf{M}$ is a (positive semi-definite) mobility tensor. This expression illustrates that, in this model, dissipative motion of a dislocation follows from elastic stress on the dislocation (the stress depends on the defect configuration).

\section{Outlook}

The model described in Sec. \ref{sec:Qtensor} has not yet been validated against the benchmark configuration involving the motion of a dislocation dipole as in Sec. \ref{sec:compat}. Whereas in the quasi-static limit of $C_{w}$ large we expect a similar evolution, this latter model has additional solutions that have not yet been investigated. In particular, there will be additional propagating modes that will mix with diffusive modes in the vicinity of the dislocation cores.

It is also apparent from the foregoing discussion, that a derivation of the phase field crystal model from a density functional theory, including stresses as a slow variable, could help resolve the issues of independence, nature of the currents, and their coupling. For example, while permeation is a dissipative mode, the stress driving it originates entirely from the reversible contribution to the stress from the free energy. Dissipation appears indirectly through the dissipative relaxation of the order parameter, but not directly through a dissipative current. The latter could, in fact, be of lower order in gradients. Along the same lines, the stress currents in Sec. \ref{sec:compat} are only dissipative until the additional compatible mode is introduced. That the reversible distortion needs to be introduced through a unphysical transformation (the displacement of the field by the compatible displacement) is difficult to justify when, in general,  reversible currents should follow from considerations of Galilean invariance within the model.

Finally, experimental advances driven by increased fluxes at Synchrotron facilities already allow the combination of near field high energy diffraction microscopy (nf-HEDM) with Bragg coherent diffractive imaging \cite{re:suter17,re:miao99,re:miao15,re:yau17}. This technique has the potential to extend the range of length scales probed by 3D X-ray diffraction techniques into the range of a few tens of nanometers, and hence to image individual lattice defects and their motion in small polycrystalline samples. These experiments will provide key information to help elucidate many of the issues regarding the phase field crystal theory discussed in this paper, and to make the technique predictive in terms of defect motion at the nanoscale.

\section*{Acknowledgments}
We are indebted to Audun Skaugen for the preparation of the figures, and to him, Vidar Skogvoll, and Marco Salvalaglio for many interesting and stimulating conversations. The research of JV has been supported by the National Science Foundation under Grant No. DMR-1838977, and the work of AA by Grant No. OIA-DMR 2021019.

\bibliographystyle{IEEEtran}
\bibliography{msmse}

\end{document}